\documentclass[prd,amsmath,amssymb,nofootinbib,preprintnumbers]{revtex4}

\usepackage{graphicx}% Include figure files
\usepackage{dcolumn}% Align table columns on decimal point
\usepackage{bm}% bold math
\usepackage{amsmath}
\usepackage{amsfonts}
\def\ls{\mathrel{\lower4pt\vbox{\lineskip=0pt\baselineskip=0pt
           \hbox{$<$}\hbox{$\sim$}}}}
\def\gs{\mathrel{\lower4pt\vbox{\lineskip=0pt\baselineskip=0pt
           \hbox{$>$}\hbox{$\sim$}}}}
\def\drawbox#1#2{\hrule height#2pt

\hbox{\vrule width#2pt height#1pt \kern#1pt
              \vrule width#2pt}
              \hrule height#2pt}

\def\Asym#1#2{\vcenter{\vbox{\drawbox{#1}{#2}
              \kern-#2pt       % line up boxes
              \drawbox{#1}{#2}}}}

\newcommand{\beq}{\begin{equation}}
\newcommand{\eeq}{\end{equation}}

\begin{document}

%
%\vspace*{2cm}
\title{Unifying inflation and dark matter}

\author{Rouzbeh Allahverdi}

\affiliation{Department of Physics \& Astronomy, University of New Mexico, Albuquerque, NM 87131, USA}

%\date{May 3, 2006}

%\maketitle

\begin{abstract}
We present a simple model where a scalar field is responsible for cosmic inflation and generates the seed for structure formation, while its thermal relic abundance explains dark matter in the universe. The inflaton self-coupling also explains the observed neutrino masses. All the virtues can be attained in a minimal extension of the Standard Model gauge group around the TeV scale. We can also unveil these properties in the forthcoming ground based experiments.
\end{abstract}

\maketitle



%%%%%%%%%%%%%%%%%%%%%%%%%%%%%%%%%%%%%%%%%%%%%%%%%%%%%%%%%%%%%%%%%%%%%%%%%%%%%%%
\section{Introduction}

The origins of inflation~\cite{Linde} and dark matter~\cite{DM} are two important problems in cosmology whose explanation requires physics beyond the electroweak standard model (SM). The two issues can be bound to the origin of neutrino masses within a minimal extension of the SM that explains all of them in a single set up~\cite{ADM}. In this model inflation occurs along a flat direction consisting of the supersymmetric partners of the right-handed (RH) neutrinos and left-handed (LH) leptons, and the Higgs field (for reviews on flat direction inflation, see~\cite{Review}). The observed anisotropy in the cosmic microwave background is related to the tiny neutrino masses. Part of the inflaton, the RH sneutrino, is stable and its thermal relic density accounts for the missing matter in the universe.

\section{The Model}

We consider the minimal supersymmetric standard model (MSSM)~\cite{Nilles} with three RH neutrino multiplets. The relevant part of the superpotential is
\begin{equation} \label{supot}
W = W_{\rm MSSM} + {\bf h} {\bf N} {\bf H}_u {\bf L}.
\end{equation}
Here ${\bf N}$, ${\bf L}$ and ${\bf H}_u$ are superfields containing the RH neutrinos, the LH leptons and the Higgs which gives
mass to the up-type quarks, respectively. For conciseness we have omitted the generation indices, and we work in a basis where neutrino Yukawa couplings $h$ (and hence neutrino masses) are diagonalized.

We note that the RH (s)neutrinos are singlets under the SM gauge group. However in many extensions of the SM they can transform non-trivially under the action of a larger gauge group. We consider a minimal extension of the SM gauge group which includes a $U(1)$: $SU(3)_c \times SU(2)_L \times U(1)_Y \times U(1)_{B-L}$, where $B$ and $L$ denote the baryon and lepton numbers respectively. This is the simplest extension of the SM symmetry which is also well motivated: anomaly cancelation requires that three RH neutrinos exist, so that they pair with LH neutrinos to form massive neutrinos.

In this model we have an extra $Z$ boson ($Z^{\prime}$) and one extra gaugino ($\tilde Z'$). The $U(1)_{B - L}$ gets broken around TeV by new Higgs fields with ${B - L} = \pm 1$. This also prohibits a Majorana mass for the RH neutrinos at the renormalizable level (note that ${\bf N N}$ has ${B - L} = 2$). The Majorana mass can be induced by a non-renormalizable operator, but it will be very small. The neutrinos will then be of Dirac nature and the value of $h$ needs to be small, i.e. $h \sim 10^{-12}$, in order to explain the light neutrino mass, $\sim {\cal O}(0.1~{\rm eV})$ corresponding to the atmospheric neutrino oscillations detected by Super-Kamiokande experiment~\cite{Super-K}.

\section{Inflation}

The ${\bf N} {\bf H}_u {\bf L}$ monomial represents a $D$-flat direction~\cite{GKM} under the $U(1)_{B-L}$, as well as the SM gauge group. The scalar field corresponding to the flat direction is denoted by
\begin{equation} \label{flat}
\phi = {{\tilde N} + H_u + {\tilde L} \over \sqrt{3}}\, ,
\end{equation}
where ${\tilde N}$, ${\tilde L}$, $H_u$ are the scalar components of corresponding superfields. The potential along the flat direction, after minimization along the angular direction, is found to be~\cite{AKM},
\begin{eqnarray} \label{flatpot}
V (\vert \phi \vert) = \frac{m^2_{\phi}}{2} \vert \phi \vert ^2 + \frac{h^2}{12} \vert \phi \vert^4 \, - \frac{A h}{6\sqrt{3}} \vert \phi \vert^3 \,.
\end{eqnarray}
where $m_\phi$ and $A$ are the soft mass and the $A$-term respectively. We note that $m_{\phi}$ is given in terms of ${\tilde N},~H_u,~{\tilde L}$ masses:
\begin{equation} \label{phimass}
m^2_{\phi} = {m^2_{\tilde N} + m^2_{H_u} + m^2_{\tilde L} \over 3}.
\end{equation}

If $A = 4 m_{\phi}$, there exists a {\it saddle point} for which $V^{\prime}(\phi_0) = V^{\prime \prime}(\phi_0) = 0$. The saddle point and the potential are given by:
\begin{eqnarray} \label{sad} \phi_0 = \sqrt{3}\frac{m_{\phi}}{h}=
6 \times 10^{12} ~ m_{\phi} ~ \Big({0.05 ~
{\rm eV} \over m_\nu} \Big)\,, \\
\label{sadpot}
V(\phi_0) = \frac{m_{\phi}^4}{4h^2}=3 \times 10^{24} ~ m^4_{\phi} ~
\Big({0.05 ~ {\rm eV} \over m_\nu} \Big)^2 \,.
\end{eqnarray}
Here $m_\nu$ denotes the neutrino mass that is given by $m_\nu = h \langle H_u \rangle$, with $\langle H_u \rangle \simeq 174$ GeV. For neutrino masses with a hierarchical pattern, the largest neutrino mass is $m_\nu \simeq 0.05$ eV in order to explain the atmospheric neutrino oscillations~\cite{Super-K}.

The slow roll inflation takes place within an interval $\Delta \phi \sim \phi^3_0/M^2_{\rm P}$ in the vicinity of $\phi_0$, and is governed by the third derivative of the potential, $V^{\prime\prime\prime}(\phi_0)=(2/\sqrt{3})hm_{\phi}$~\cite{AEGM,AEGJM}~\footnote{Initial condition for inflation can be naturally set within prior phase(s) of false vacuum inflation either by the help of quantum fluctuations~\cite{AFM}, or as a result of attractor behavior of the potential at $\phi_0$~\cite{ADM2}. Many models of high energy physics possess metastable vacua, and hence can lead to false vacuum in°ation at some stage during the evolution of the early universe.}. A sufficient number of e-foldings ${\cal N}_e \sim 10^3$ is generated during the slow roll regime~\cite{AKM}. The amplitude of density perturbations follows~\cite{AKM}.
\begin{equation} \label{amp}
\delta_{H} \simeq \frac{1}{5\pi}\frac{H^2_{\rm inf}}{\dot\phi} \simeq 3.5 \times 10^{-27}
~ \Big( {m_\nu \over 0.05 ~ {\rm eV}} \Big)^2 ~  \Big({M_{\rm P}
\over m_{\phi}} \Big) ~ {\cal N}_{\rm COBE}^2\,,
\end{equation}
where $H_{\rm inf}$ denotes the Hubble rate during inflation which is given by $H^2_{\rm inf} = 8 V(\phi_0)/3 M^2_{\rm P}$. Here $m_{\phi}$ denotes the loop-corrected value of the inflaton mass at the scale $\phi_0$ in Eqs.~(\ref{sad},\ref{sadpot},\ref{amp}). In Figure 1, we show the neutrino mass as a function of the inflaton mass. We use $\delta_H = 1.91 \times 10^{-5}$ to draw the plot. We see that the neutrino mass in the range $0$ to $0.30$~eV corresponds to the inflaton mass of $0$ to $30$~ GeV.

Moreover, a scalar spectral index within the $2 \sigma$ range allowed by 5-year WMAP data $0.934 \leq n_s \leq 0.988$~\cite{WMAP5} can be obtained if $\phi_0$ becomes a point of inflection, i.e. $V^{\prime}(\phi_0) \neq 0,~V^{\prime \prime}(\phi_0) = 0$ ~\cite{LYTH1,AEGJM}. This happens for a slight departure from a saddle point $\Big(\vert A^2 - 16 m^2_{\phi} \vert^{1/2}/4m_{\phi}\Big) \ls 10^{-8}$~\cite{AEGJM}. Note that its reflection on the amplitude of the density perturbations is negligible.

\begin{figure}[t]
\includegraphics[width=7cm]{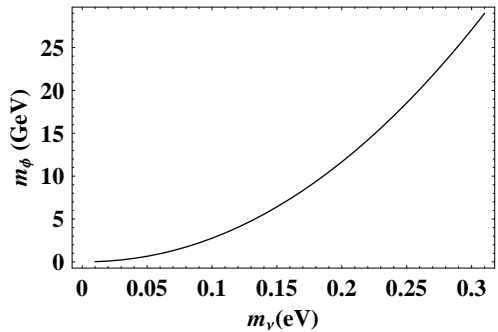}
\caption{The inflaton mass $m_{\phi}$ is plotted as a function of
the neutrino mass $m_{\nu}$.} \label{neutrinoinf}
\end{figure}

\section{Dark Matter}

The inflaton has gauge couplings to the electroweak and $U(1)_{\rm B-L}$ gauge/gaugino fields. It therefore induces a VEV-dependent mass $\sim g \langle \phi \rangle$ for these fields ($g$ denotes a typical gauge coupling). After the end of inflation, $\phi$ starts oscillating around the global minimum at the origin with a frequency $m_{\phi} \sim 10^3 H_{\rm inf}$. When the inflaton passes through the minimum, $\langle \phi \rangle = 0$, the induced mass undergoes non-adiabatic time variation. This results in non-perturbative particle production~\cite{PREHEAT1,PREHEAT2}. As the inflaton VEV is rolling back to its maximum value $\phi_0$, the mass of the gauge/gaugino quanta increases again. Because of their large couplings they quickly decay to the fields which are not coupled to the inflaton, hence massless, notably the down-type (s)quarks. This is a very efficient process as a result of which relativistic particles are created within few Hubble times after the end of inflation (for more details, see~\cite{AEGJM}). A thermal bath of MSSM particles is eventually formed with a temperature $T_{\rm R} \sim 10^6$ GeV.

Since the RH sneutrino $\tilde N$ is a singlet under the SM gauge group, its mass receives the smallest contribution from quantum corrections due to gauge interactions. It can therefore be set to be the lightest supersymmetric particle (LSP), hence the dark matter candidate. Note that part of the inflaton, i.e. its ${\tilde N}$ component see Eq.~(\ref{flat}), has never decayed; only the coherence in the original condensate that drives inflation is lost. Its relic abundance will be then set by thermal freeze-out.

The mass of dark matter (i.e. RH sneutrino) is correlated with the inflaton mass, see Eq.~(\ref{phimass}). However, the former is calculated at the weak scale, while the latter is at the scale $\phi_0$, see Eq.~(\ref{sad}). The two quantities are related to each other by renormalization group equations (RGEs). In order to calculate the masses in the model, we use SUGRA boundary conditions (for details, see~\cite{ADM})~\footnote{A similar analysis in the context of MSSM inflaiton has been done for neutralino dark matter~\cite{ADM3}.}. Even though the inflaton mass is small at the scale $\phi_0$, obtained by solving Eq.~(\ref{sad}), the RGE effects increase the RH sneutrino and slepton masses at the weak scale. In Figure 2 we plot the RH sneutrino and stop masses at the weak scale for different values of gluino masses. The gluino masses for the three lines (solid and dashed) are 730 GeV, 1.20 TeV and 1.64 TeV (bottom-up). The lines are drawn for 0.30 eV neutrino mass (but do not shift much for a neutrino mass of 0.05 eV). The masses of sparticles (e.g., the lighter stop) can be within the reach of initial phase of the LHC.

\begin{figure}[t]
\includegraphics[width=7cm]{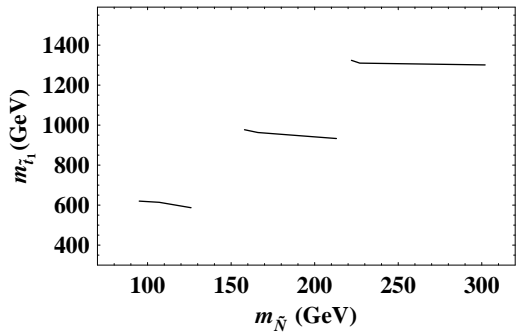}
\caption{$m_{\tilde t_1}$ vs $m_{\tilde N}$. The lines are for
neutrino masses 0.3 eV. The gluino masses for the three lines are
730, 1200 and 1640 GeV (bottom-up). For smaller neutrino masses, the
lines are slightly shifted downwards.} \label{sneutrinoinf}
\end{figure}

The primary diagrams responsible to provide the right amount of relic density are mediated by ${\tilde Z}^{\prime}$ in the $t$-channel. In Figure 3, we show the relic density values for smaller masses of sneutrino where the lighter stop mass is $ \ls 1$ TeV (hence easily accessible at the LHC). We have varied new gaugino and Higgsino masses and the ratio of the vacuum expectation values (VEVs) of new Higgs fields to generate Fig. 3. We find that the WMAP allowed values of the relic density are satisfied for many points.

\begin{figure}[t]
\includegraphics[width=7cm]{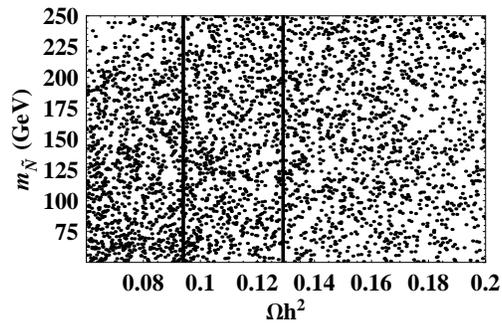}
\caption{$\Omega h^2$ vs $m_{\tilde N}$. The solid lines from left
to right are for $\Omega h^2 =$ 0.094 and 0.129 respectively.
}\label{sneutrinoominf}
\end{figure}

Since the RH sneutrino interacts with quarks via the $Z^{\prime}$ boson, it is possible to see it via the direct detection experiments. The detection cross sections are not small as the interaction diagram involves $Z^{\prime}$ in the $t$-channel. The typical cross section is about 2$\times 10^{-8}$ pb for a $Z^{\prime}$ mass around 2 TeV. It is therefore possible to probe this model in the upcoming dark matter detection experiments~\cite{Direct}. The signal for this new scenario at the LHC will contain standard jets plus missing energy and jets plus leptons plus missing energy. The jets and the leptons will be produced from the cascade decays of squarks and gluinos into the final state containing the sneutrino. In general, the models with RH sneutrino dark matter will have similar signal as the models with neutralino dark matter. The major difference, however, lies in the fact that the lightest dark matter particle has different spin, i.e., that sneutrino has spin 0 compared to the lightest neutralino whose spin is 1/2.

\section{Acknowledgments}

The author is indebted to Bhaskar Dutta, Alex Kusenko and Anupam Mazumdar for collaboration and numerous discussions on various aspects of the physics presented here.


\begin{thebibliography}{99}

\bibitem{Linde}
For a review, see: A. D. Linde, hep-th/0503203.

\bibitem{DM}
For example, see: G. Bertone, D. Hooper and J. Silk, Phys. Rept. {\bf 405}, 279 (2005), hep-ph/0404175.

\bibitem{ADM}
R.~Allahverdi, B.~Dutta and A.~Mazumdar, Phys. Rev. Lett. {\bf 99}, 261301 (2007), arXiv:0708.3983 [hep-ph].

\bibitem{Review}
R. Allahverdi, Mod. Phys. Lett. A {\bf 23}, 2799 (2008), arXiv:0812.3628 [hep-ph].

\bibitem{Nilles}
For a review on supersymmetry, see: H. P. Nilles, Phys. Rept. {\bf 110}, 1 (1984).

\bibitem{Super-K}
Y. Fukuda {\it et al.} [Super-Kamiokande Collaboration], Phys. Rev. Lett. {\bf 81}, 1562 (1998), hep-ex/9807003.

\bibitem{GKM}
T.~Gherghetta, C.~F.~Kolda and S.~P.~Martin, Nucl.\ Phys.\ B {\bf 468}, 37 (1996), hep-ph/9510370.

\bibitem{AKM}
R.~Allahverdi, A.~Kusenko and A.~Mazumdar, JCAP {\bf 0707}, 018 (2007), hep-ph/0608138.

\bibitem{AEGM}
R.~Allahverdi, K.~Enqvist, J.~Garcia-Bellido and A.~Mazumdar, Phys. Rev. Lett. {\bf 97}, 191304 (2006), hep-ph/0605035.

\bibitem{AEGJM}
R.~Allahverdi, K.~Enqvist, J.~Garcia-Bellido, A.~Jokinen and A.~Mazumdar, JCAP {\bf 0706}, 019 (2007), hep-ph/0610134.

\bibitem{AFM}
R.~Allahverdi, A.~R.~Frey and A.~Mazumdar, Phys. Rev. D {\bf 76}, 026001 (2007), hep-th/0701233.

\bibitem{ADM2}
R. Allahverdi, B. Dutta and A. Mazumdar, Phys. Rev. D {\bf 78}, 063507 (2008), arXiv:0806.4557 [hep-ph].

\bibitem{WMAP5}
E.~Komatsu {\it et al.} [WMAP Collaboration], arXiv:0803.0547 [astro-ph].

\bibitem{LYTH1}
J.~C.~Bueno Sanchez, K.~Dimopoulos and D.~H.~Lyth, JCAP {\bf 0701}, 015 (2007), hep-ph/0608299.

\bibitem{PREHEAT1}
L.~Kofman, A.~D.~Linde and A.~A.~Starobinsky, Phys.\ Rev.\ Lett.\ {\bf 73}, 3195 (1994), hep-th/9405187.

\bibitem{PREHEAT2}
L.~Kofman, A.~D.~Linde and A.~A.~Starobinsky, Phys.\ Rev.\ D {\bf 56}, 3258 (1997), hep-ph/9704452.

%\bibitem{EM}
%K.~Enqvist and A.~Mazumdar, Phys. Rept. {\bf 380}, 99 (2003).

%\bibitem{DK}
%M.~Dine and A.~Kusenko, Rev.\ Mod.\ Phys.\  {\bf 76}, 1 (2004).

\bibitem{ADM3}
R.~Allahverdi, B.~Dutta and A.~Mazumdar, Phys. Rev. D {\bf 75}, 075018 (2007), hep-ph/0702112.

\bibitem{Direct}
L. Baudis, arXiv:0711.3788 [astro-ph].


\end{thebibliography}
\end{document}